\documentclass[aps,twocolumn,showpacs,floatfix]{revtex4-2}
\usepackage{dcolumn,multirow}
\usepackage{graphicx}
\usepackage{amsfonts,mathtools,amsmath,bm,amssymb}
\usepackage{comment}
\usepackage[usenames]{color}
\usepackage[flushleft]{threeparttable}
\usepackage{titlesec}
\usepackage{etoolbox} 
\usepackage{lipsum} 
\usepackage[capitalize]{cleveref}
\usepackage{makecell}
\usepackage[section]{placeins}
\usepackage{float}
\usepackage{color}

\usepackage[T1]{fontenc}
\usepackage{textcomp}
\usepackage[utf8]{inputenc}
\DeclareUnicodeCharacter{00B2}{\ensuremath{{}^2}}

\begin{document}
	\title{ Superconducting topological Dirac semimetals: $P6/m$-Si$_6$ and $P6/m$-NaSi$_6$  }
	\author{Alex Taekyung Lee} \email{neotechlee@gmail.com}
	\affiliation{Department of Applied Physics, Yale University, New Haven, Connecticut 06520, USA}
	\author{Kyungwha Park} 
	\affiliation{Department of Physics, Virginia Tech, Blacksburg, Virginia 24061, USA}
	\author{In-Ho Lee}  \email[corresponding author ]{ihlee@kriss.re.kr}
	\affiliation{Korea Research Institute of Standards and Science, Daejeon 34113, Korea}
	\date{\today }

	\begin{abstract}
		We theoretically propose that hexagonal silicon-based crystals, $P6/m$-Si$_6$ and $P6/m$-NaSi$_6$, are topological Dirac semimetals with superconducting critical temperatures of 12 K and 13 K, respectively, at ambient pressure. 
		Band inversion occurs with the Fu-Kane topological invariant $\mathbb{Z}_2=1$, even in the absence of spin-orbit coupling. 
		The Dirac nodes protected by $C_6$ crystal rotational symmetry remain gapless with spin-orbit coupling. Using first-principles calculations, we find pressure-induced topological phase transitions for $P6/m$-Si$_6$ and $P6/m$-NaSi$_6$ with critical external pressures of 11.5 GPa and 14.9 GPa, respectively. 
		Above the critical pressures, the Dirac bands are gapped with
		$\mathbb{Z}_2=0$, while the superconducting states and the crystal symmetries are retained. 
		Our results may shed light into a search for silicon-based topological materials with superconductivity.
	\end{abstract}
	
	\maketitle
	
	\section{Introduction}
	Semiconducting silicon becomes indispensable in electronics due to its versatile features such as the ease of electron or hole doping in a wide range, high-temperature stability, non toxicity, and natural abundance. 
	It is not uncommon to modify phases of solids by varying crystal structures and/or applying external stimuli. 
	There has been a great effort in fabricating silicon in different condensed matter phases, especially
	a superconducting phase. 
	Some metallic silicon phases were proposed to be superconducting under high pressures \cite{BuckelPL1965,Stepanov1980,Martinez1986,ErskinePRL1986,ChangPRL1985}. For doped silicon clathrates and boron-doped cubic silicon, superconductivity was observed at ambient pressure \cite{Kawaji1995,Connetable2003,Bustarret2006}. 
	Recently, hexagonal silicon-based crystals $P6/m$-Si$_6$ and $P6/m$-NaSi$_6$, 
	which can be synthesized similarly to Ref.~\onlinecite{KimNatMat2015}, 
	were predicted to show superconductivity (transition temperature of 12 K and 13 K, respectively) 
	at ambient pressure \cite{HJSungPRL2018}.

	Topology in the reciprocal space plays an important role in quantum materials properties due to topological protection. 
	Based on symmetries, gapped and gapless quantum phases including a superconducting phase can be topologically classified \cite{Chiu2016}. 
	For example, three-dimensional Dirac semimetals which can exist in the presence of time-reversal and inversion symmetries, are categorized into two classes \cite{YangNcomm2014}. 
	In the first class, band inversion occurs with a nontrivial Fu-Kane $\mathbb{Z}_2$ topological invariant \cite{Fu2007} and associated Fermi arc surface states, and Dirac nodes are protected by crystal rotational symmetry \cite{ZWang2012,WangPRB2013}. 
	This class is henceforth referred to as a topological Dirac semimetal phase, following Ref.~\cite{YangNcomm2014}. 
	In the second class, Dirac nodes are protected by nonsymmorphic space group \cite{YoungPRL2012} without band inversion.

	Most topological materials are not based on silicon. 
	A topological Dirac semimetal phase was discovered in a silicon-containing compound, CaAl$_2$Si$_2$ (space group No. 164, point group $D_{3d}$) \cite{Hao_Su2020,Tao_Deng2020} in the presence of spin-orbit coupling (SOC). 
	Silicon-based crystals $Cmcm$-AHT-Si$_{24}$ and $Cmcm$-VFI-Si$_{36}$ 
    have been proposed to be topological nodal line semimetals \emph{only} 
	when SOC is excluded \cite{LiuJPCL2019}. 
	With SOC, the nodal lines are gapped with a small energy gap of 1 meV \cite{LiuJPCL2019}.

	Recently, superconductivity was observed in three-dimensional topological Dirac semimetals such as Cd$_3$As$_2$\cite{AggarwalNmat2016,WangNmat2016,Huangncomm2019}, Au$_2$Pb\cite{SchoopPRB2015,XingNPJ2016}, KZnBi\cite{SongPRX2021}, PdTe$_2$\cite{PTeknowijoyoPRB2018,AmitPRB2018}, and BaSn$_3$\cite{HuangPRB2021}.
	To the best of our knowledge, silicon-based topological Dirac semimetals with superconductivity 
	have not been proposed or synthesized yet. 
	Superconductivity in topological Dirac semimetals is appealing considering the possibility 
	to create odd-parity or spin-triplet Cooper pair interaction 
	by doping and/or breaking time-reversal symmetry \cite{KobayashiPRL2015,Hashimoto2016,Sato2017}. 
	Furthermore, superconductors with a nontrivial Fu-Kane $\mathbb{Z}_2$ topological invariant were proposed to be
	platforms for realization of Majorana zero modes \cite{Alicea2012} 
	whose braiding is used for topological quantum computation \cite{Kitaev2003,Sarma2015}. 
	Families of superconductors with a nontrivial $\mathbb{Z}_2$ invariant showed promising experimental
	signatures of Majorana zero modes at the ends of magnetic vortices \cite{Kreisel2020,Kong2021}.
	
	
	\begin{figure*}[tb!]
		\begin{center}
			\includegraphics[width=0.98\textwidth, angle=0]{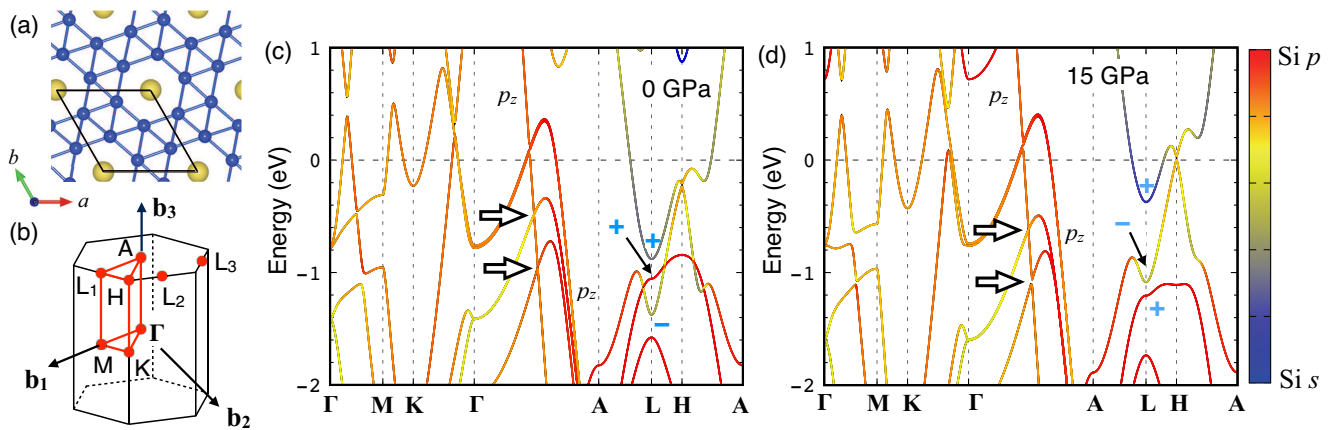}
			\caption{(a) Top view of the atomic structure of $P6/m$-NaSi$_6$ with a unit cell.
				Isostructure $P6/m$-Si$_6$ can be obtained by removing Na atoms from $P6/m$-NaSi$_6$.
				(b) First Brillouin zone for $P6/m$-Si$_6$ and $P6/m$-NaSi$_6$.
				SOC-included DFT band structures of $P6/m$-Si$_6$ at (c) 0 GPa and (d) 15 GPa, where the Fermi level is set to zero.
				All bands are at least doubly degenerate. Relative proportions of Si $s$ and $p$-orbital characters 
				are encoded in colors.
				The Dirac points are indicated as arrows and the parity of a few occupied bands are shown.
			}
			\label{atm_str_band}
		\end{center}
	\end{figure*}

	In this work, 
	we propose that $P6/m$-Si$_6$ and $P6/m$-NaSi$_6$ are topological Dirac semimetals with superconductivity 
	at ambient pressure even when SOC is included. 
	We find that the silicon-based crystals undergo a topological phase transition 
	driven by hydrostatic pressure without any structural changes. 
	Above slightly different critical pressures, both silicon-based crystals 
	become topologically trivial in the superconducting state.
	

	\section{Methods}
	We employ density functional theory (DFT) with projector augmented wave (PAW)~\cite{PAW} pseudopotentials
	and the Perdew-Burke-Ernzerhof generalized gradient approximation (PBE-GGA) \cite{PBE}
	for the exchange-correlation functional, as implemented in the \textsf{VASP} software~\cite{VASP}.
	A plane-wave basis set with a kinetic energy cutoff of 500 eV is used.
	We use $\Gamma$-centered \textbf{k}-point meshes of $8 \times 8 \times 20$ for bulk $P6/m$-Si$_6$ and $P6/m$-NaSi$_6$ and $8 \times 8 \times 1$ for finite slabs. 
	Starting with the geometries from Ref.~\onlinecite{HJSungPRL2018}, 
	we further optimize the atomic coordinates and lattice constants
	until the residual forces are less than $0.01$ eV/{\AA} with and without pressure. 
	The optimized bulk hexagonal lattice parameters for $P6/m$-Si$_6$ ($P6/m$-NaSi$_6$) are $a=6.81$ (6.76) {\AA} and $c= 2.50$ (2.44) {\AA} without pressure, which are close to those in Ref.~\onlinecite{HJSungPRL2018}. 
	The relaxed atomic coordinates can be found in Appendix \ref{appendix:atm_str}.

	We compute the Fu-Kane $\mathbb{Z}_2$ topological invariant \cite{Fu2007} 
	from the parity of all valence or occupied bands at the time-reversal invariant momentum (TRIM) points
	as well as by using the Wilson loop method \cite{SoluyanovPRB2011,WU2017}.
	To apply the Wilson loop method and to investigate the surface states, 
	we first construct a tight-binding Hamiltonian for the Si $s+p$
	bands by generating 24 (33) maximally localized Wannier functions for $P6/m$-Si$_6$ ($P6/m$-NaSi$_6$), 
	using \textsf{WANNIER90} code \cite{MLWF,wannier20} (see Appendix \ref{appendix:Z2}).
	Then, we calculate the surface Green's function of the semi-infinite system 
	based on the Wannier-function tight-binding model, using \textsf{WannierTools} code\cite{WU2017}. 
	Irreducible representations of bands at the high symmetry \textbf{k} points and along the $\Gamma$-A axis 
	are obtained using \textsf{Quantum Espresso}~\cite{QE-2017}.

	\section{Results and discussion}
	The atomic structure of $P6/m$-NaSi$_6$ (space group no. 175, point group $C_{6h}$)  
	is shown in Fig.~\ref{atm_str_band}(a), 
	where the stacking along the $c$ axis is identical for all atomic layers. 
	The structure of $P6/m$-Si$_6$ is obtained by simply removing Na atoms from that of $P6/m$-NaSi$_6$. (see Appendix \ref{appendix:atm_str}.)
	The Na atoms can be easily removed by the degassing process, since the migration barrier of Na atoms along the cylindrical holes is only $0.48$ eV~\cite{HJSungPRL2018}. 
	This barrier is $0.17$ eV lower than that for the $Cmcm$ phase~\cite{KimNatMat2015}.
	

	\begin{figure}
		\begin{center}
			\includegraphics[width=0.45\textwidth, angle=0]{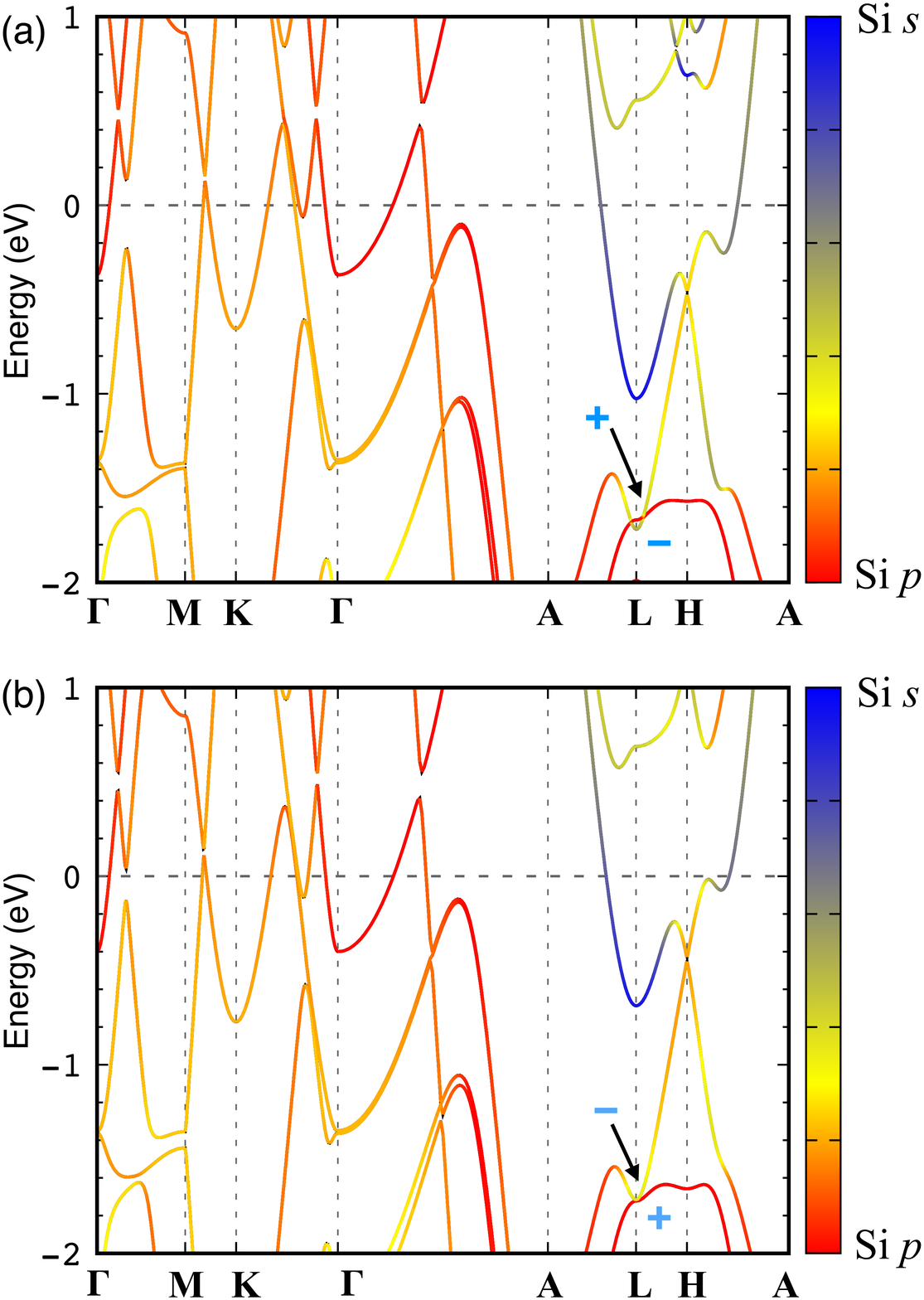}
			\caption{Electronic band structure obtained from first-principles calculations using DFT.
				First-principles band structures of $P6/m$-NaSi$_6$ at pressure (a) 0 GPa and (b) 15 GPa,
				including spin-orbit coupling. According to the relative weight of the Si $s$ character and the Si $p$ character, the band color was encoded as shown in the legend. 
				The band structure calculation at pressure 15 GPa shows the band structure just before band inversion occurs. The situation of band inversion that occurs with changes in pressure can be seen, especially in the L-H section.
			}
			\label{NaSi6_bands}
		\end{center}
	\end{figure}
	
	Let us first discuss electronic and topological properties without pressure and then with pressure.
	Fig.~\ref{atm_str_band}(c) shows the electronic structure of $P6/m$-Si$_6$ at symmetric \textbf{k} points with SOC in the absence of pressure.
	Mostly $p$-orbital character is dominant near the Fermi level. 
	However, $s$-orbital character is locally dominant in the $k_z=\pi/c$ plane, \emph{i.e.}, along the A-L-H-A lines. 
	At the L point, bands at $-0.86$ eV and $-1.04$ eV (relative to the Fermi level) have strong $s$-orbital and $p_z$-orbital characters, respectively, 
	while bands at $-1.36$ eV have characters of mixed $s$ and $p_x+p_y$ orbitals.
	(see Appendix \ref{appendix:band_str}.)
	Strong Si $p$-orbital characteristics also appear in most of the \textbf{k} space 
	except for the $k_z=\pi/c$ plane for $P6/m$-NaSi$_6$,
	as presented in Fig.~\ref{NaSi6_bands}.

	\begin{table}
		\begin{ruledtabular}
			\begin{center}
				\caption{ Irreducible representation (irrep), point group symmetries, 
					and their eigenvalues of bands at L point, for $P6/m$-Si$_6$ at 0 GPa.
					$C_2$ is two-fold rotation operator with respect to the $z$-axis.
					$I$ and $\sigma_h$ are two-fold inversion and horizontal mirror plane operators, respectively. 
					The highest occupied band is 24th band. }  \label{symm_at_L}
				\begin{tabular}{c c c c c}
					band \# &	irrep		& $C_2$	& $\sigma_h$	& $I$ 		\\
					\hline
					1, 2	& $B_u$	& $-1$	& $+1$	& $-1$	\\
					3, 4	& $A_g$	& $+1$	& $+1$	& $+1$	\\
					5, 6	& $B_g$	& $-1$	& $-1$	& $+1$	\\
					7, 8	& $B_u$	& $-1$	& $+1$	& $-1$	\\
					9, 10	& $A_u$	& $+1$	& $-1$	& $-1$	\\
					11, 12	& $A_g$	& $+1$	& $+1$	& $+1$	\\
					13, 14	& $B_g$	& $-1$	& $-1$	& $+1$	\\
					15, 16	& $A_u$	& $+1$	& $-1$	& $-1$	\\
					17, 18	& $B_u$	& $-1$	& $+1$	& $-1$	\\
					19, 20	& $A_g$	& $+1$	& $+1$	& $+1$	\\
					21, 22	& $A_u$	& $+1$	& $-1$	& $-1$	\\
					23, 24	& $B_u$	& $-1$	& $+1$	& $-1$	\\
					\hline
					25, 26	& $B_g$	& $-1$	& $-1$	& $+1$	\\
				\end{tabular}
			\end{center}
		\end{ruledtabular}
	\end{table}

	For $P6/m$-Si$_6$ at 0 GPa, the parity analysis at the TRIM points gives $\mathbb{Z}_2=1$ 
	(see Appendix \ref{appendix:Z2}.) 
	which results from the band inversion at the L points, 
	as shown in Fig.~\ref{atm_str_band}(c). (Note that a unit cell of $P6/m$-Si$_6$ has 24 valence electrons.)
	At the L points, the 23th and 24th (25th and 26th) bands at $-1.36$~eV ($-1.04$~eV) have $-$ ($+$) parity, 
	as depicted in Fig.~\ref{atm_str_band}(c). 
	The band inversion is not induced by SOC because $\mathbb{Z}_2=1$ without SOC.

	The analysis of irreducible representations at the L point indicates that 
	the opposite parity for the two band pairs originates from the horizontal mirror symmetry
	$\sigma_h$. 
	The irreducible representation and eigenvalues of $\sigma_h$ of the bands at the L point are listed in Table \ref{symm_at_L}.
	The 23th/24th bands belong to irreducible representation $B_u$ 
	(where an eigenvalue of horizontal mirror reflection $\sigma_h$ is $+1$), 
	while the 25th/26th bands belong to irreducible representation $B_g$ (where an eigenvalue of $\sigma_h$ is $-1$). 
	We also confirm that $\mathbb{Z}_2=1$ using the Wilson loop method. 
	(see Appendix \ref{appendix:Z2}.) 
	Even when the parity of all occupied bands below the Fermi
	level is counted, we find that $\mathbb{Z}_2=1$ persists.

	\begin{figure}[htb!]
		\begin{center}
			\includegraphics[width=0.48\textwidth, angle=0]{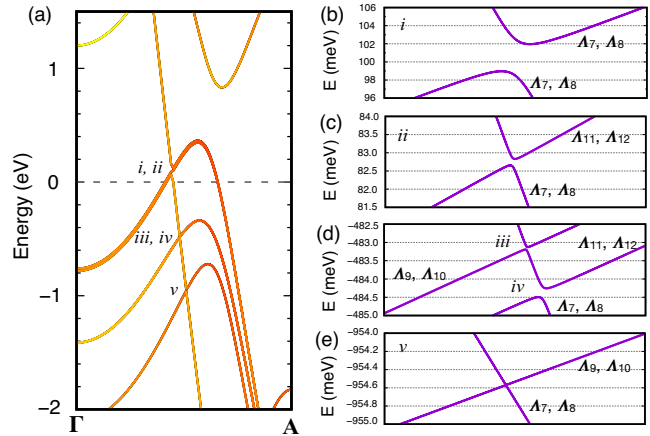}
			\caption{SOC-included DFT bands for (a) $P6/m$-Si$_6$ at 0 GPa, between $\Gamma$ and A. (b)-(e) enlarged bands near 5 possible band crossing points ($i$-$v$) for (a).  $\Lambda_i$ represent irreducible representations of double group $C_6$ along the $\Gamma$-A direction. 
				Points $iii$ and $v$ are gapless Dirac nodes (when $\Lambda_{7,8}$ bands and $\Lambda_{9,10}$ 
				bands meet each other).}
			\label{Si6_dirac}
		\end{center}
	\end{figure}

	For $P6/m$-Si$_6$ with SOC in the absence of pressure, along the $\Gamma$-A direction, 
	we find several possible band crossing points in the vicinity of the Fermi level. 
	To identify gapless Dirac nodes, we zoom-in the bands near the crossing points with numerical accuracy of 10~$\mu$eV, as shown in Fig.~\ref{Si6_dirac}. 
	The bands at $i$, $ii$, and $iv$ are gapped with 0.07-28.9 meV, 
	while the crossing points $iii$ and $v$ are gapless within numerical accuracy. 
	We also analyze irreducible representations $\Lambda_i$ of the bands near the crossing points, 
	considering that double group $C_6$ is applied to the bands along the $\Gamma$-A direction. 
	This symmetry analysis [Figs.~\ref{Si6_dirac}(b)-(e)] agrees with the numerical result. 
	The Dirac nodes along the $k_z$ axis are protected by the crystal sixfold rotational symmetry $C_6$. 
	The gapless protected Dirac nodes in conjunction with our analysis of the $\mathbb{Z}_2$ invariant, 
	suggest that $P6/m$-Si$_6$ is a topological Dirac semimetal.

\begin{figure}[tb!]
		\begin{center}
			\includegraphics[width=0.5\textwidth, angle=0]{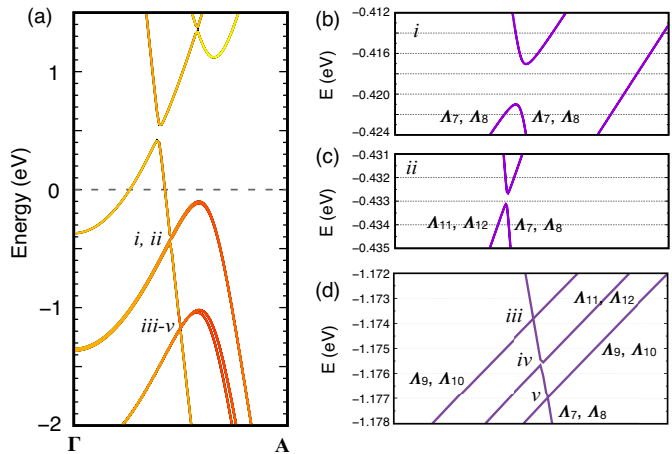}
			\caption{ (a) DFT bands in $P6/m$-NaSi$_6$ at 0 GPa, between $\Gamma$ and A. 
				(b)-(d) enlarged bands near 5 possible band crossing points ($i$-$v$) for (a). 
				Points $iii$ and $v$ are gapless Dirac nodes (when $\Lambda_{7,8}$ bands and $\Lambda_{9,10}$  
				bands meet each other).
			} 
			\label{NaSi6_dirac}
		\end{center}
	\end{figure}

	For $P6/m$-NaSi$_6$, it is tricky to compute the $\mathbb{Z}_2$ invariant by counting the parity of $N$ bands 
	at the TRIM points, 
	where $N$ is the number of valence electrons, 
	since $N$ is now an odd number due to the Na atom. (In Ref.~\onlinecite{Fu2007}, each degenerate pair was counted only
	once for centrosymmetric metals.) 
	In order to circumvent this, we take into account all occupied bands below the Fermi level at the TRIM points 
	in our calculation of the $\mathbb{Z}_2$ invariant. 
	Note that there are different numbers of occupied bands at different TRIM points. 
	For $P6/m$-NaSi$_6$ at 0 GPa, the product of the parity values of all occupied bands at each L point is positive, while the corresponding products at the other 5 TRIM points are negative. 
	(See  Appendix \ref{appendix:Z2}.) 
	This gives rise to $\mathbb{Z}_2=1$ and the band inversion also occurs at the L points,
	as shown in Fig.~\ref{NaSi6_bands}(a).
	Therefore, similarly to $P6/m$-Si$_6$, $P6/m$-NaSi$_6$ is also a topological Dirac semimetal. 
	The gapless Dirac nodes [crossing points $iii$ and $v$ in Fig.~\ref{NaSi6_dirac}(d)] 
	are found below the Fermi level.
	
\begin{figure}[tb!]
		\begin{center}
			\includegraphics[width=0.48\textwidth, angle=0]{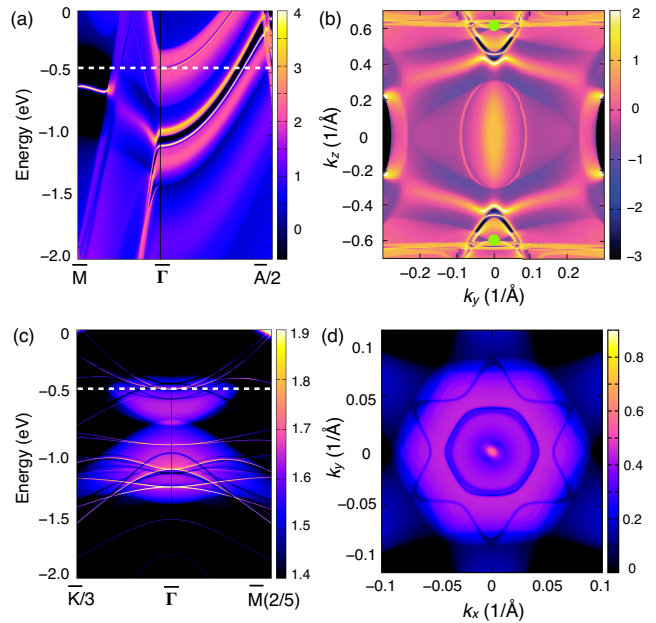}
			\caption{Electronic structure or local density of states of surface states along the two-dimensional symmetric \textbf{k} points (a,c), in the $k_y$-$k_z$ plane (b) and in the $k_x$-$k_y$ plane (d) for the semi-infinite (100) and (001) surfaces of $P6/m$-Si$_6$, respectively. 
				Horizontal dashed lines in (a) and (c) represent the Dirac point energies, respectively.
				Filled dots in (b) represent the projection points of the Dirac nodes onto the $k_y$-$k_z$ plane.
				In (a)-(d), bright color indicates higher density of states at the surface.}
			\label{Si6_arc}
		\end{center}
	\end{figure}

	The nontrivial topology of three-dimensional Dirac bands in $P6/m$-Si$_6$ and $P6/m$-NaSi$_6$ indicates that there are nontrivial surface states. 
	We calculate surface states using the surface Green's function of the semi-infinite system described earlier, considering two surface types: surfaces 
	parallel and perpendicular to the $\Gamma$-A direction, $i.e.$, (100) and (001) surfaces, respectively.
	Fig.~\ref{Si6_arc} shows the calculated local density of states which is an imaginary part of the surface Green's function, for the (100) and (001) surfaces. 
	At the chemical potential, corresponding to one of the Dirac node energies, for the (100) surface, 
	two arc-shaped (left and right sides of hourglass) surface states connect the two Dirac node projection points (with bulk characteristics) indicated as solid dots along the $k_z$ axis in Fig.~\ref{Si6_arc}(b). 
	This is expected because each Dirac node consists of degenerate Weyl nodes with opposite chirality in topological Dirac semimetals \cite{ZWang2012,Villanova2017}.
	For the (001) surface, the surface states appear as a point [Fig.~\ref{Si6_arc}(d)]. 
	In this case, the two Dirac nodes along the $k_z$ axis are projected onto the same point, $i.e.$, the origin, in the $k_x$-$k_y$ plane. 
	Thus, point-like surface states are expected rather than arc-shaped surface states. 
	The features of the surface states for the (100) and (001) surfaces agree with those of prototype topological Dirac semimetals Na$3$Bi\cite{ZWang2012} and Cd$_3$As$_2$\cite{WangPRB2013}.

	Now with pressure up to 15 GPa, the $P6/m$ crystal symmetry is still maintained for $P6/m$-Si$_6$ and $P6/m$-NaSi$_6$. 
	Since the effect of pressure is similar for both crystals, 
	as shown in Figs. \ref{atm_str_band} and \ref{NaSi6_bands},
	we mainly discuss $P6/m$-Si$_6$.
	At 15 GPa, the amount of changes of the band structure depends on orbital characters. 
	Specifically, the bands of $s$-orbital character are sensitive to pressure, 
	while the bands of $p$-orbital character are somewhat rigid [Fig.~\ref{atm_str_band}(d)]. 
	The band structure near the Fermi level changes significantly only in the $k_z=\pi/c$ plane [Fig.~\ref{atm_str_band}(c) vs (d)]. 
	At the L points, the $s$-orbital band is shifted by $0.51$ eV, 
	whereas the highest-energy occupied $p$-orbital band
	is shifted by 0.16 eV.

\begin{figure}[tb!]
		\begin{center}
			\includegraphics[width=0.48\textwidth, angle=0]{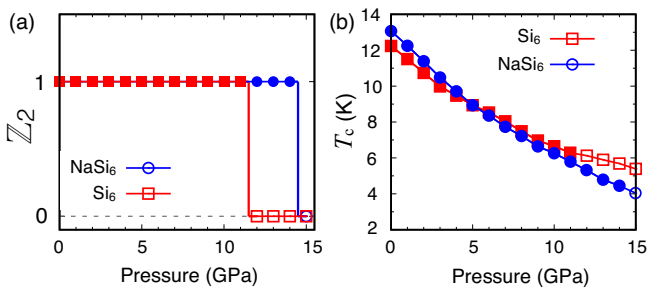}
			\caption{ (a) $\mathbb{Z}_2$ topological invariants for $P6/m$-Si$_6$ and $P6/m$-NaSi$_6$ are shown as a function of external hydrostatic pressure.
				Critical external pressures of the topological phase transitions for $P6/m$-Si$_6$ and $P6/m$-NaSi$_6$ are $11.5$~GPa and $14.9$~GPa, respectively.
				(b) Superconducting critical temperatures $T_c$ of $P6/m$-Si$_6$ and $P6/m$-NaSi$_6$, obtained from Ref.~\onlinecite{HJSungPRL2018} are shown as a function of external pressure. In (a) and (b), the filled (empty) symbols indicate the $\mathbb{Z}_2=1$ ($\mathbb{Z}_2=0$) phase.}
			\label{z2_pressure}
		\end{center}
	\end{figure}

	While the topological invariant is insensitive to small local perturbations such as disorder or impurities, pressure-induced deformation can be used to control the electronic structure \cite{YoungPRB2011,Bahramy2012,Bansil2016,IdeuePNAS2019}.
	For $P6/m$-Si$_6$, under high pressure, 
	the parity of the 23th/24th bands is reversed to that of the 25th/26th bands at the L point
	[See the bands in the energy window between $-1.4$ and $-1.0$ eV in Fig.~\ref{atm_str_band}(d)], 
	and so the bands with $-$ parity become higher in energy than the bands with $+$ parity, 
	which removes the band inversion. 
	This leads to a topological phase transition from nontrivial $\mathbb{Z}_{2}=1$ to trivial $\mathbb{Z}_{2}=0$. 
	The critical external pressure for the topological phase transition in $P6/m$-Si$_6$ is 11.5 GPa. 
	[See Fig.~\ref{z2_pressure}(a).] 
	The gapless Dirac nodes which exist below the critical pressure now open up a gap above the critical pressure, 
	as shown in Fig.~\ref{atm_str_band}(d), due to the topological phase transition. 
	In the case of $P6/m-$NaSi$_6$, a similar topological phase transition occurs at 14.9 GPa, 
	which is somewhat higher than the critical pressure of $P6/m$-Si$_6$.

	Theoretically, $P6/m$-Si$_6$ and $P6/m$-NaSi$_6$ 
	were proposed to be superconductors at ambient pressure\cite{HJSungPRL2018}.
	Fig.~\ref{z2_pressure}(b) shows the superconducting critical temperature $T_c$ 
	as a function of hydrostatic pressure. 
	While $T_c$ decreases monotonically as pressure ($P$) increases, 
	both $P6/m$-Si$_6$ and $P6/m$-NaSi$_6$ 
	are superconducting within the pressure range we consider, $0 \leq P \leq 15$ GPa. 
	Therefore, we propose that $P6/m$-Si$_6$ ($P6/m$-NaSi$_6$) 
	is a superconducting topological Dirac semimetal below 11.5 (14.9) GPa. 
	We emphasize that $P6/m$-Si$_6$ and $P6/m$-NaSi$_6$ consist of light elements with atomic number 14 or less, 
	based on silicon, in contrast to reported superconducting topological Dirac semimetals such as Cd$_3$As$_2$\cite{AggarwalNmat2016,WangNmat2016,Huangncomm2019}, Au$_2$Pb\cite{SchoopPRB2015}, KZnBi\cite{SongPRX2021}, PdTe$_2$\cite{PTeknowijoyoPRB2018,AmitPRB2018}, and BaSn$_3$\cite{HuangPRB2021} 
	which consist of heavy elements.


	
	
	
	Due to the nesting vectors along $k_z$, it turns out that there is a modulation
	in Si-Si bond lengths (out-of-hexagonal plane) formed in the $z$ direction.
	Indeed, we found a small oscillation of the out-of-plane bond length in the slab geometry. (see Appendix \ref{appendix:bond_mod}.)
	The modulation is strongest on the surface and decreases near the center of the slab.

\begin{figure}[tb!]
		\begin{center}
			\includegraphics[width=0.48\textwidth, angle=0]{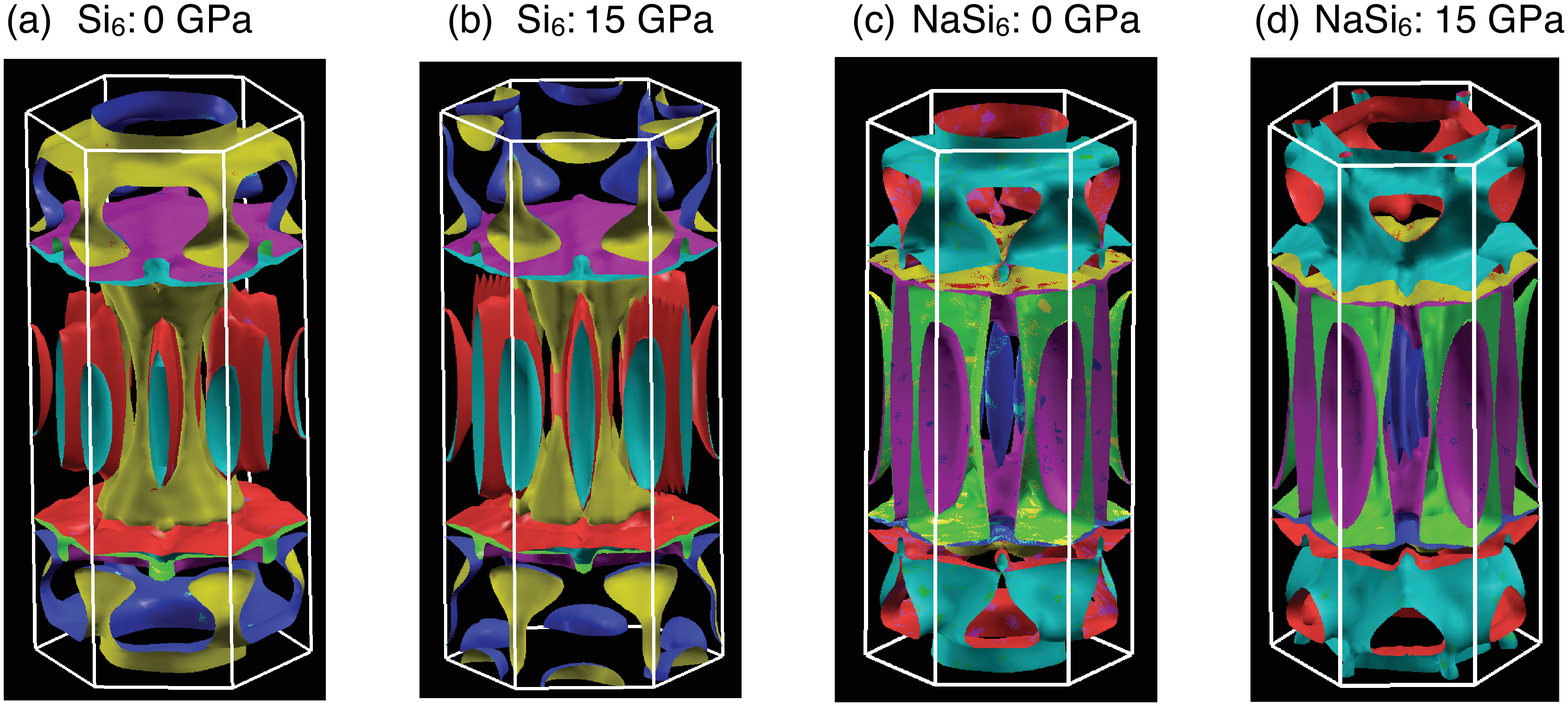}
			\caption{Fermi surfaces of (a) $P6/m$-Si$_6$ at 0 GPa, (b) $P6/m$-Si$_6$ at 15 GPa,
				(c) $P6/m$-NaSi$_6$ at 0 GPa, and (d) $P6/m$-NaSi$_6$ at 150 GPa
				are shown, respectively. 
				The Fermi surface is the surface of constant energy in the first BZ 
				which separates occupied from unoccupied electron states at zero temperature. Electronic band structure obtained from first-principles calculations using DFT.
			}
			\label{fermisurf2}
		\end{center}
	\end{figure}

	We calculated the Fermi surfaces from the two crystal structures, $P6/m$-Si$_6$ and $P6/m$-NaSi$_6$, as presented in Fig. \ref{fermisurf2}.
	$P6/m$-Si$_6$ and $P6/m$-NaSi$_6$ can provide Dirac cones near or at the Fermi energy
	and show an anisotropic conducting channel due to their anisotropic bonding nature.
	Evidence of our proposed pressure-induced topological phase transition in $P6/m$-Si$_6$ and $P6/m$-NaSi$_6$ 
	may be explored by transport experiments. 
	The three-dimensional Dirac nodes and resultant arc-shaped surface states are supposed to induce a unique field-direction dependence in Shubnikov-de Haas oscillations \cite{Moll2016}. 
	The chiral anomaly associated with Weyl nodes suggests the following interesting properties: 
	(i) A large negative magnetoresistance is found when an external magnetic field is parallel to a current direction \cite{Burkov2011,JXiong2015}; 
	(ii) Thermoelectric properties depends on the relative direction between an external magnetic field and a temperature gradient \cite{ZJia-NComm-2016}; 
	(iii) The giant planar Hall effect is expected when a current direction is not parallel to a magnetic field direction in plane \cite{Burkov2017,MWu2018}. 
	By measuring variations of the above properties upon external pressure, the proposed topological phase transition
	can be experimentally probed, similarly to Ref.~\onlinecite{IdeuePNAS2019}.

	\section{Conclusions}
	
	In summary, we have proposed that $P6/m$-Si$_6$ and $P6/m$-NaSi$_6$ crystals 
	are superconducting topological Dirac semimetals at ambient pressure  with $\mathbb{Z}_2=1$. 
	The two gapless bulk Dirac nodes appear at $-0.4832$ eV and $-0.9546$ eV for the former and at $-1.1740$ eV 
	and $-1.1770$ eV for the latter below the Fermi level. 
	With hole doping, the Fermi level may be lowered to one of the two Dirac node energies 
	for experimental signatures of the topological phase. 
	The gapless Dirac nodes are protected by the crystal rotational symmetry in the presence of time-reversal symmetry. 
	The topological Dirac semimetal phase for the crystals becomes topologically trivial 
	beyond a critical external pressure. 
	This pressure-induced topological phase transition retains superconductivity 
	and the original crystal symmetry group.
	The coexistence of a topological Dirac semimetal state and a superconducting state in the present crystals, 
	along with a pressure-induced topological phase transition, 
	will provide an interesting platform to study the interplay between topology 
	in electronic structure and superconductivity.

	%
	%

	~\\
	\section{Acknowledgments}
	We thank Sohrab Ismail-Beigi for helpful discussions.
	This work also used the Extreme Science and Engineering Discovery Environment (XSEDE),
	which is supported by the National Science Foundation grant number ACI-1548562, by using computer time 
	on the Comet supercomputer as enabled by XSEDE allocation MCA08X007. 
 I.H.L was supported by the National Center for Materials Research Data (NCMRD) through the National Research Foundation of Korea (NRF) funded by the Ministry of Science and ICT (NRF-2021M3A7C2089748). 
	
	\bibliography{myref}

	
	\appendix
	
	\section{Atomic structure of $P6/m$-S\lowercase{i}$_6$}
	\label{appendix:atm_str}

\begin{figure}
		\begin{center}
			\includegraphics[width=0.45\textwidth, angle=0]{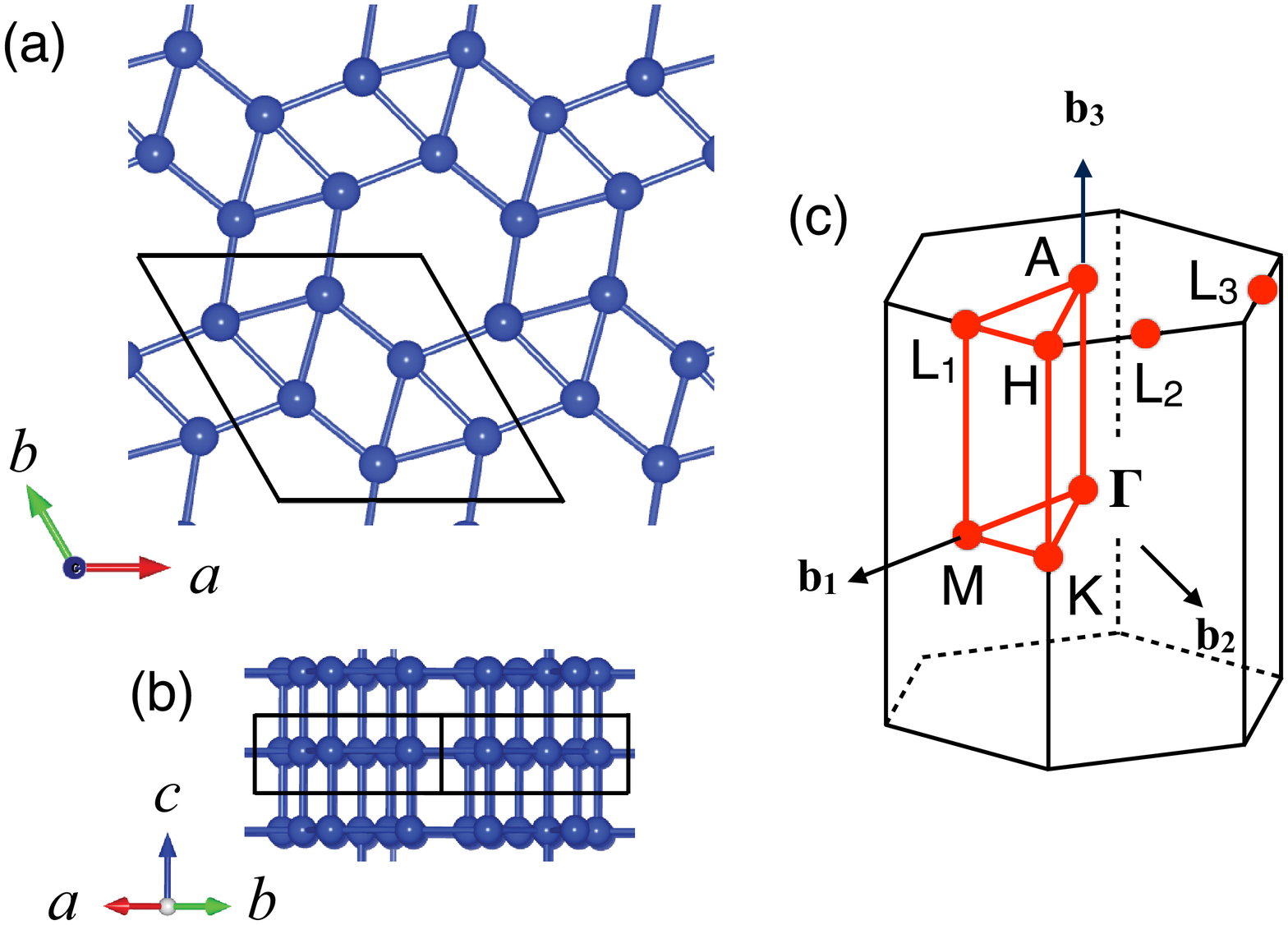}
			\caption{(a) Top view and (b) side view of the atomic structure of $P6/m$-Si$_6$.
				(c) First Brillouin zone (BZ) of $P6/m$-Si$_6$. 
				$\Gamma$, M, A, and L points are time reversal invariant momentum points, 
				which are related to $\mathbb{Z}_2$ topological invariant calculation
				especially for crystals with inversion symmetry. 
				$P6/m$ (space group no. 175) is belong to the hexagonal Bravais lattice crystal system. The space group no. 175 has inversion symmetry.
			}
			\label{atm_str2}
		\end{center}
	\end{figure}

Fig. \ref{atm_str2} shows the atomic structure of $P6/m$-Si$_6$, which can be obtained by 
removing Na atoms from $P6/m$-NaSi$_6$.
In Table \ref{lat_param}, we provide the information of the atomic structure including lattice parameters
and atomic coordinate. 
Note that all the structures have $P6/m$ space group symmetry, at both 0 GPa and 15 GPa.

	\begin{table*}
		\begin{ruledtabular}
			\begin{center}
				\caption{Optimized lattice parameters and atomic coordinates of $P6/m$-Si$_6$ and $P6/m$-NaSi$_6$ 
					phases at external pressure ($P$) of 0 GPa and 15 GPa,
					obtained from first-principles calculations using density-functional theory (DFT).
					Both $P6/m$-NaSi$_6$ and $P6/m$-Si$_6$ phases are hexagonal with $\alpha = \beta=90^\circ$, and $\gamma=120^\circ$. }
				\label{lat_param}
				\begin{tabular}{c c c c c c c c}
					Material 			& $P$ (GPa)	& $a=b$ (\AA)	& $c$ (\AA)	& Atom	& $x$	& $y$	& $z$	 \\
					\hline
					$P6/m$-Si$_6$		& 0	    	& 6.813	& 2.501	& Si ($6j$)		& 0.154	& 0.733	& 0	\\
					$P6/m$-Si$_6$		& 15		& 6.542	& 2.402	& Si ($6j$)		& 0.145	& 0.725	& 0	\\
					$P6/m$-NaSi$_6$	& 0	    	& 6.755	& 2.444	& Na ($1b$)	& 0	& 0	& 0.5		\\
					& 	    	&		&		& Si ($6j$)		& 0.145	& 0.710	& 0	\\
					$P6/m$-NaSi$_6$	& 15		& 6.536	& 2.377	& Na ($1b$)	& 0	& 0	& 0.5		\\
					& 	    	&		&		& Si ($6j$)		& 0.144	& 0.713	& 0	\\
				\end{tabular}
			\end{center}
		\end{ruledtabular}
	\end{table*}

	\section{$\mathbb{Z}_2$ topological invariant calculations}
	\label{appendix:Z2}

\begin{figure}
		\begin{center}
			\includegraphics[width=0.45\textwidth, angle=0]{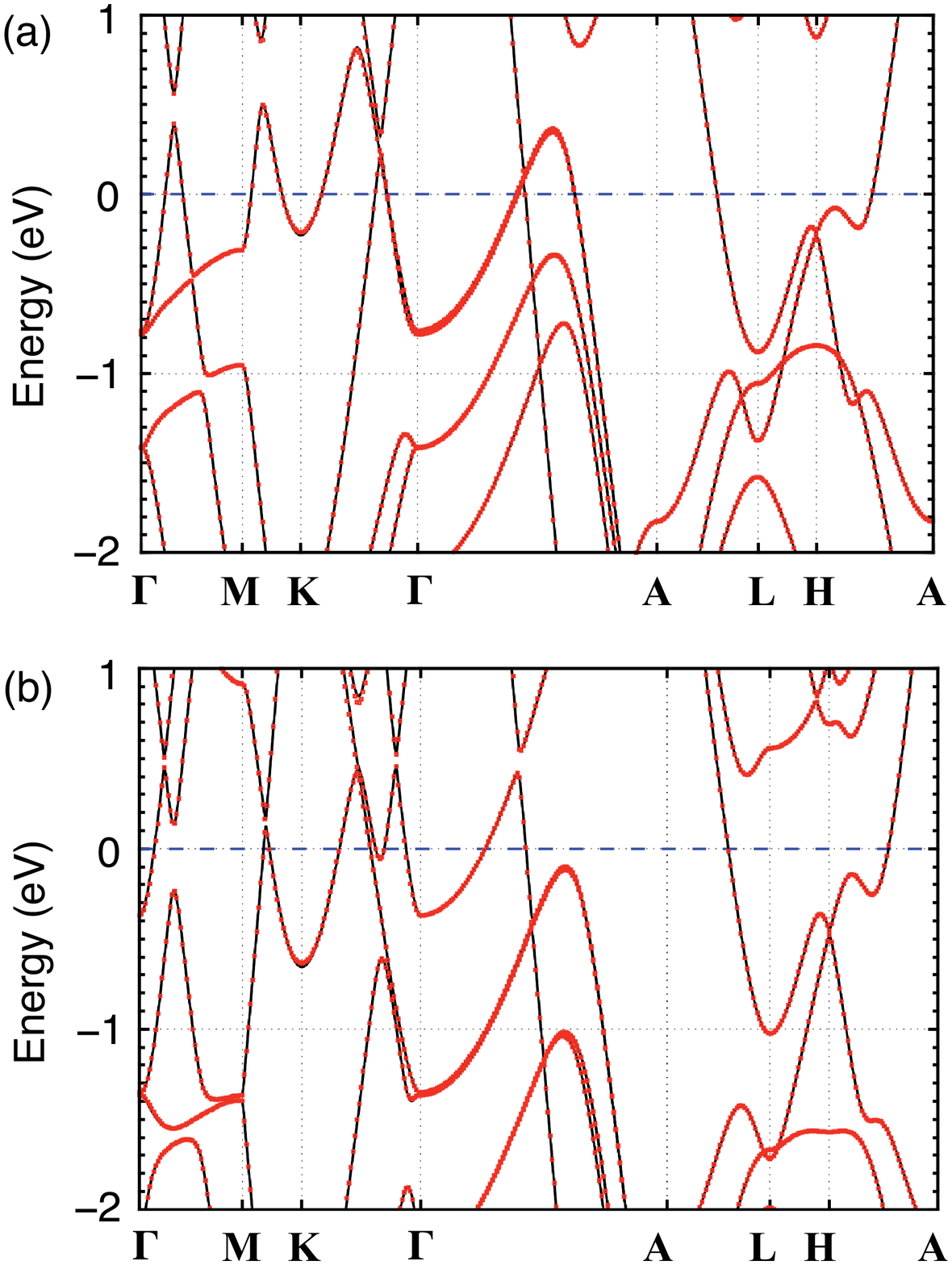}
			\caption{ First-principles bands (solid line) 
				and Wannier bands (red filled circles) 
				of $P6/m$-Si$_6$ at  (a) 0 GPa and (b) 15 GPa.
				In the eV energy scale, there are no visible differences between two different bands.  
				Spin-orbit coupling is included. 
			}
			\label{wann_bands}
		\end{center}
	\end{figure}

 	The topological phase of band structure can be classified by calculating topological invariant.
We check the $Z_2$ topology by calculating the parities at time-reversal invariant momenta, 
as shown in Table \ref{parities}, and using the hybrid Wannier charge center flow technique, which is equivalent to the Wilson loop method.
	To investigate the topological nature of the electronic states, 
	we obtained maximally localized Wannier functions using the \textsf{WANNIER90} code \cite{MLWF,wannier20} (see Fig. \ref{wann_bands})
	and performed hybrid Wannier charge center flow calculations.
	With this Wannier model, 
	the symmetry properties and $\mathbb{Z}_2$ topological invariant could be directly computed. 
	$\mathbb{Z}_2$ topological invariants were calculated by observing hybrid Wannier charge center flows. The topological surface states were calculated using the Green's function approach, 
	as implemented in the \textsf{WannierTools} code \cite{WU2017}.

	\begin{table}
		\begin{ruledtabular}
			\begin{center}
				\caption{ Parities at time-reversal invariant momentum points of $P6/m$-Si$_6$ and $P6/m$-NaSi$_6$ at 0 GPa and 15 GPa.
				}
				\label{parities}
				\begin{tabular}{c c c c c c}
					&		& \multicolumn{2}{c}{$P6/m$-Si$_6$}	& \multicolumn{2}{c}{$P6/m$-NaSi$_6$} \\
					\cline{3-4} \cline{5-6}
					& {\bf k}-point	& 0 GPa	      & 15 GPa	 & 0 GPa	& 15 GPa \\
					\hline
					\rm{$\Gamma$}	& (0,0,0)	      & $+$	& $+$		& $-$	& $-$	\\
					\rm{M$_1$}		& (0.5, 0, 0)	  & $+$	& $+$		& $-$	& $-$	\\
					\rm{M$_2$}		& (0, 0.5, 0)	  & $+$	& $+$		& $-$	& $-$	\\
					\rm{A}			& (0, 0, 0.5)	  & $+$	& $+$		& $-$	& $-$	\\
					\rm{M$_3$}		& (0.5, 0.5, 0)	  & $+$	& $+$		& $-$	& $-$	\\
					\rm{L$_1$}		& (0.5, 0, 0.5)	  & $-$	& $+$		& $+$	& $-$	\\
					\rm{L$_2$}		& (0, 0.5, 0.5)	  & $-$	& $+$		& $+$	& $-$	\\
					\rm{L$_3$}		& (0.5, 0.5, 0.5) & $-$	& $+$		& $+$	& $-$	\\
				\end{tabular}
			\end{center}
		\end{ruledtabular}
	\end{table}

	Space group no. 175 ($P6/m$, hexagonal) has inversion symmetry.
	Because of the existence of inversion symmetry in both $P6/m$-Si$_6$ and $P6/m$-NaSi$_6$, 
	the Fu-Kane parity criterion can be used to easily calculate the $\mathbb{Z}_2$ topological invariants.
	We were able to obtain the same $\mathbb{Z}_2$ topological invariants for both $P6/m$-Si$_6$ and $P6/m$-NaSi$_6$
	by using two different methods, 
	observing hybrid Wannier charge center flows 
	and investigating Fu-Kane parity criterion.
	$\nu_0$ is the strong topological index. 
	$\nu^{\prime}_n$ ($n=1, 2, 3$) are weak topological indices.
	Table~\ref{nu_si6} show the example for $P6/m$-Si$_6$ at 0 GPa.
	From the hybrid Wannier charge center flow calculations, we calculated $\nu^{\prime}_n$ ($n=1,2,3$).
	$\nu_0$ is obtained by 
	\begin{equation} \tag{S1}
		\nu_0 = \nu_1+\nu^{\prime}_1 = \nu_2+ \nu^{\prime}_2 = \nu_3+ \nu^{\prime}_3 ~~~{\rm mod ~2} ~,
	\end{equation}
	and thus we have $(\nu_0; \nu^{\prime}_1 \nu^{\prime}_2 \nu^{\prime}_3) = (1; 0 0 1)$.
	Therefore, $P6/m$-Si$_6$ at 0 GPa has a strong topological nature.
	Similarly, we find that $P6/m$-NaSi$_6$ at 0 GPa has a strong topological nature.
	

	\begin{table}
		\begin{ruledtabular}
			\begin{center}
				\caption{ $\nu_0$ and $\nu^{\prime}_n$ ($n=1, 2, 3$) for $P6/m$-Si$_6$
					obtained from hybrid Wannier charge center flow calculations, 
					and correspond $\{k_n\}$ points and {\bf k}-planes. 
				}
				\label{nu_si6}
				\begin{tabular}{c c c c}
					$k_n$ & $k$-plane & $\nu$ \\
					\hline
					$k_1$=0.0	& $k_2$-$k_3$ plane	& $\nu_1$=1	\\
					$k_1$=0.5	& $k_2$-$k_3$ plane	& $\nu^{\prime}_1$=0	\\
					$k_2$=0.0	& $k_1$-$k_3$ plane	& $\nu_2$=1	\\
					$k_2$=0.5	& $k_1$-$k_3$ plane	& $\nu^{\prime}_2$=0	\\
					$k_3$=0.0	& $k_1$-$k_2$ plane	& $\nu_3$=0	\\
					$k_3$=0.5	& $k_1$-$k_2$ plane	& $\nu^{\prime}_3$=1	\\
				\end{tabular}
			\end{center}
		\end{ruledtabular}
	\end{table}

	\section{Band structures of $P6/m$-NaSi$_6$ and $P6/m$-Si$_6$ }
	\label{appendix:band_str}

\begin{figure}
		\begin{center}
			\includegraphics[width=0.48\textwidth, angle=0]{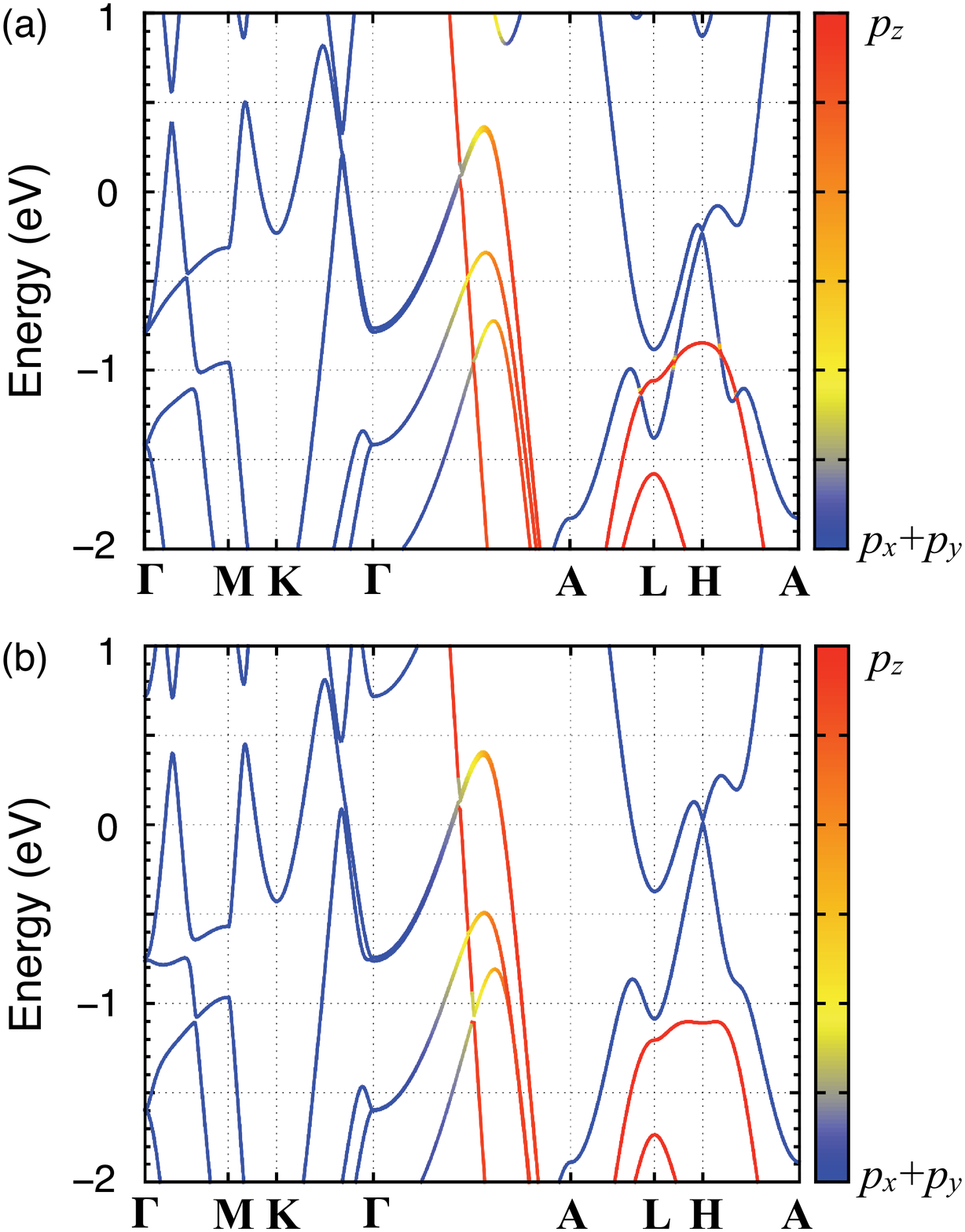}
			\caption{Electronic band structure obtained from first-principles calculations using DFT.
				In the presence of SOC, band structures of $P6/m$-Si$_6$ at (a) 0 GPa and (b) 15 GPa
				and their projections on the the Si $p$ character are shown.
				According to the relative weight of in-plane Si $p$ character ($p_x + p_y$) 
				and out-of-plane Si $p$ character ($p_z$)
				the band colors were encoded as shown in the legend. 
				The situation of band inversion that occurs with changes in external pressure ($> 11.5$ GPa) 
				can be seen,  especially in the L-H-A interval.
			}
			\label{Si6_p}
		\end{center}
	\end{figure}

 In Fig. \ref{Si6_p}, we present the the weight of in-plane Si $p$ character ($p_x + p_y$) 
and out-of-plane Si $p$ character ($p_z$) on the band structure of $P6/m$-Si$_6$.
At 0 GPa, $p_z$-dominant band is higher in energy than the $p_x + p_y$ band along the L-H-A interval, 
whereas the $p_x + p_y$ band has lower energy than the $p_z$ band at 15 GPa.

\section{ S\lowercase{i}-S\lowercase{i} bond lengths in the slab geometry }
\label{appendix:bond_mod}

\begin{figure}
		\begin{center}
			\includegraphics[width=0.40\textwidth, angle=0]{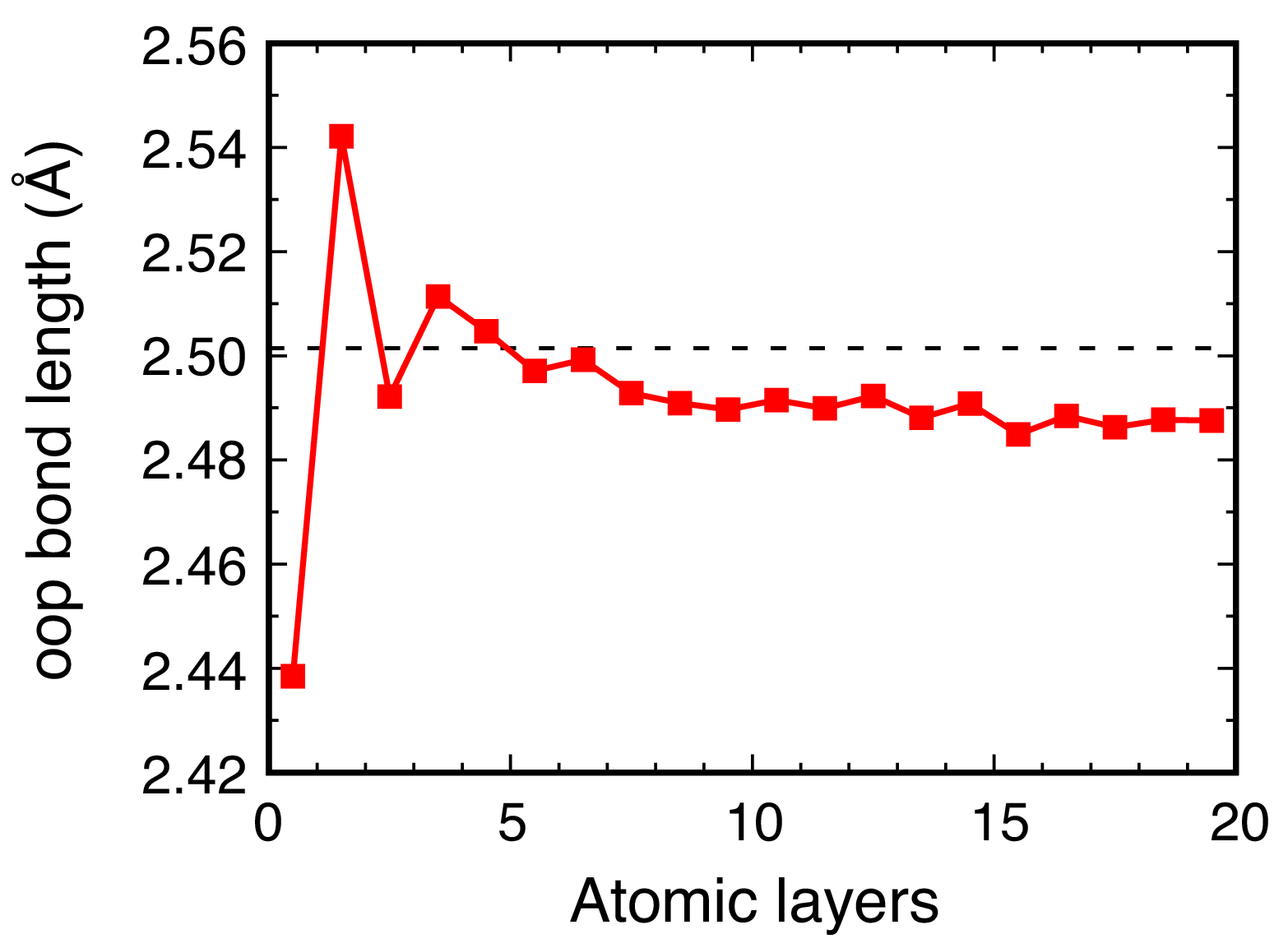}
			\caption{ Si-Si bond lengths, formed in the $z$ direction 
				also denoted by out-of-plane (oop bond lengths) from the hexagonal plane, 
				in the relaxed 40-layer $P6/m$-Si$_6$ slab are shown. 
				Position of the surface along the out-of-plane, is set to be zero. 
				Dashed line represents the Si-Si bond length in the unrelaxed slab.
			}
			\label{bond_mod}
		\end{center}
	\end{figure}

In Fig. \ref{bond_mod}, we plot the Si-Si bond length along the $z$ direction 
(out of plane from the hexagonal surface).
Note that the bond length distribution has an oscillation, while the oscillation of the bond length 
is largest and decreases near the bulk-like region. 
 
\end{document}


\title{{\Large Supplemental Material} \\
Superconducting Dirac semimetals: $P6/m$-Si$_6$ and $P6/m$-NaSi$_6$   } 

\author{Alex Taekyung Lee} \email{neotechlee@gmail.com}
\affiliation{Department of Applied Physics, Yale University, New Haven, Connecticut 06520, USA}
\author{Kyungwha Park} \email{kyungwha@vt.edu}
\affiliation{Department of Physics, Virginia Tech, Blacksburg, Virginia 24060, USA}
\author{In-Ho Lee}  \email[corresponding author ]{ihlee@kriss.re.kr}
\affiliation{Korea Research Institute of Standards and Science, Daejeon 34113, Korea}
\date{\today }
\maketitle
\tableofcontents

\section{Atomic structure of $P6/m$-S\lowercase{i}$_6$}
\label{sec:atomic_structure}

\begin{figure}[ht!] 
\begin{center}
\includegraphics[width=0.60\textwidth, angle=0]{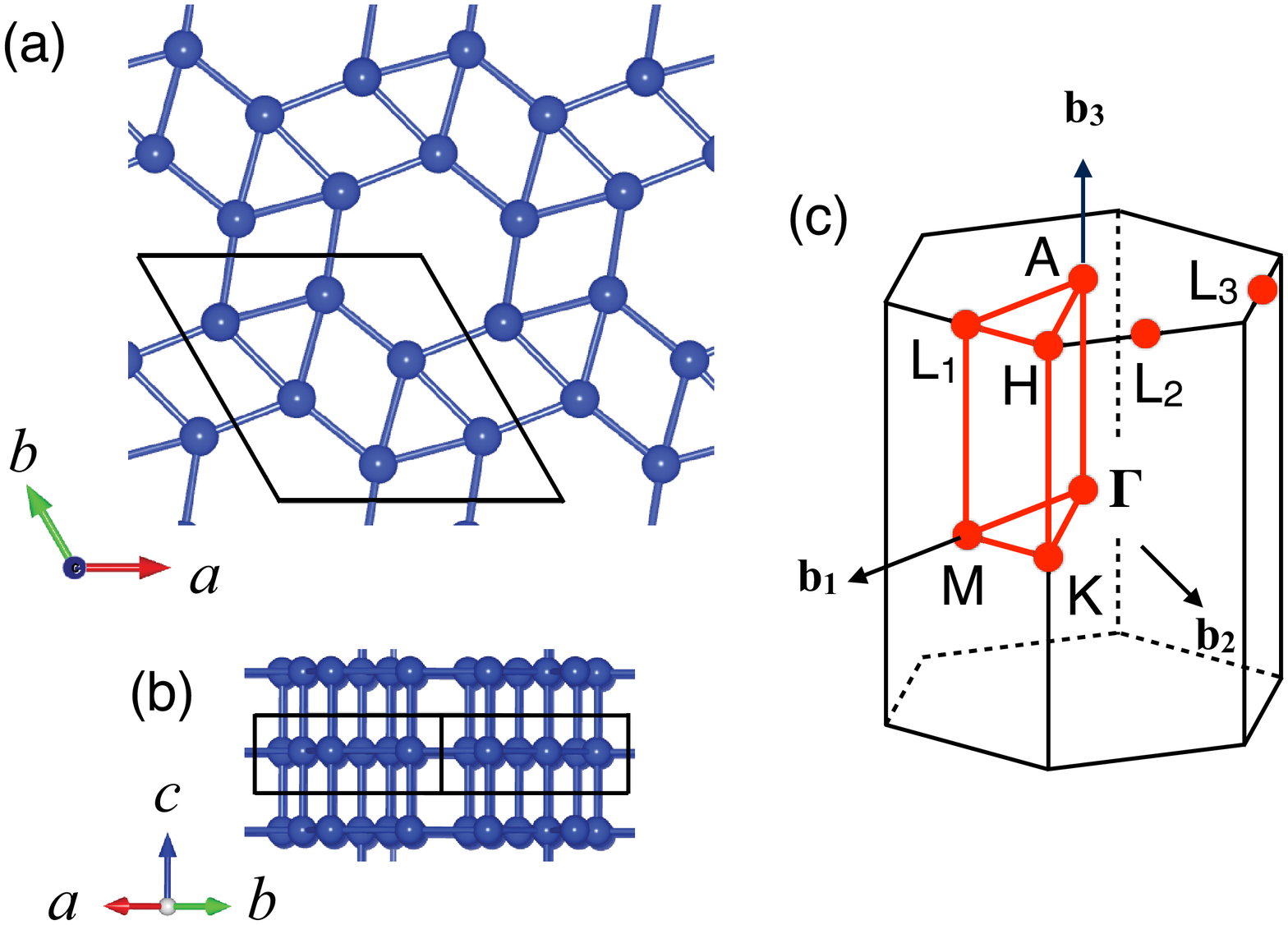}
\caption{(a) Top view and (b) side view of the atomic structure of $P6/m$-Si$_6$.
$P6/m$-Si$_6$ can be obtained by removing Na atoms from $P6/m$-NaSi$_6$.
(c) First Brillouin zone (BZ) of $P6/m$-Si$_6$. 
$\Gamma$, M, A, and L points are time reversal invariant momentum points, 
which are related to $\mathbb{Z}_2$ topological invariant calculation
especially for crystals with inversion symmetry. 
$P6/m$ (space group no. 175) is belong to the hexagonal Bravais lattice crystal system. The space group no. 175 has inversion symmetry.
}
\label{atm_str2}
\end{center}
\end{figure}

 
\begin{table}
\begin{ruledtabular}
\begin{center}
\caption{Optimized lattice parameters and atomic coordinates of $P6/m$-Si$_6$ and $P6/m$-NaSi$_6$ 
phases at external pressure of 0 GPa and 15 GPa,
obtained from first-principles calculations using density-functional theory (DFT).
Both $P6/m$-NaSi$_6$ and $P6/m$-Si$_6$ phases are hexagonal with $\alpha = \beta=90^\circ$, and $\gamma=120^\circ$. }
\label{lat_param}
\begin{tabular}{c c c c c c c c}
Material 			& pressure (GPa)	& $a=b$ (\AA)	& $c$ (\AA)	& Atom	& $x$	& $y$	& $z$	 \\
\hline
$P6/m$-Si$_6$		& 0	    	& 6.813	& 2.501	& Si ($6j$)		& 0.154	& 0.733	& 0	\\
$P6/m$-Si$_6$		& 15		& 6.542	& 2.402	& Si ($6j$)		& 0.145	& 0.725	& 0	\\
$P6/m$-NaSi$_6$	& 0	    	& 6.755	& 2.444	& Na ($1b$)	& 0	& 0	& 0.5		\\
				& 	    	&		&		& Si ($6j$)		& 0.145	& 0.710	& 0	\\
$P6/m$-NaSi$_6$	& 15		& 6.536	& 2.377	& Na ($1b$)	& 0	& 0	& 0.5		\\
				& 	    	&		&		& Si ($6j$)		& 0.144	& 0.713	& 0	\\
\end{tabular}
\end{center}
\end{ruledtabular}
\end{table}

\section{Band structures of 
$P6/m$-N\lowercase{a}S\lowercase{i}$_6$
and
$P6/m$-S\lowercase{i}$_6$ 
}

\begin{figure}[!ht]
\begin{center}
\includegraphics[width=0.7\textwidth, angle=0]{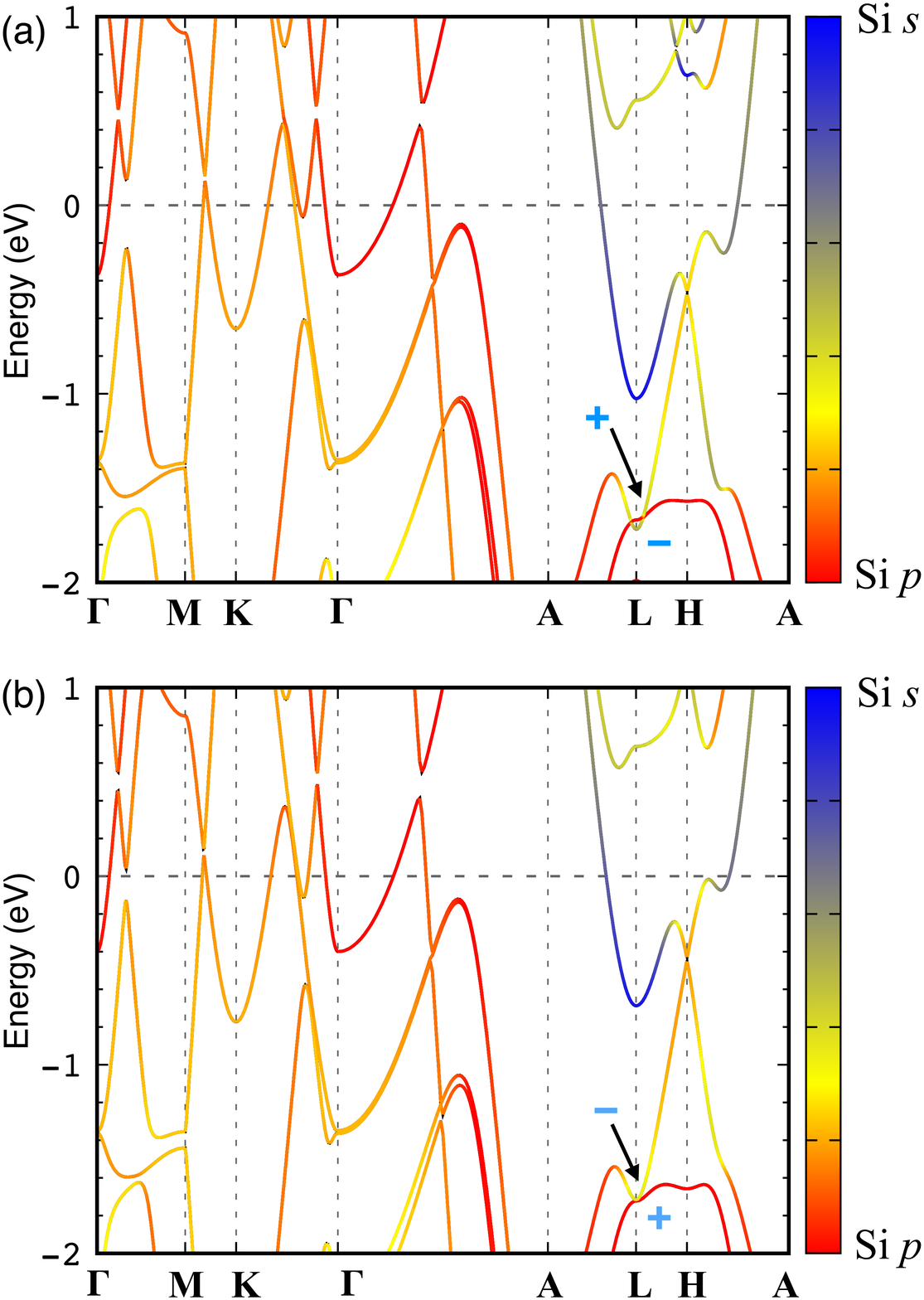}
\caption{Electronic band structure obtained from first-principles calculations using DFT.
First-principles band structures of $P6/m$-NaSi$_6$ at pressure (a) 0 GPa and (b) 15 GPa,
including spin-orbit coupling. According to the relative weight of the Si $s$ character and the Si $p$ character, the band color was encoded as shown in the legend. 
The band structure calculation at pressure 15 GPa shows the band structure just before band inversion occurs. The situation of band inversion that occurs with changes in pressure can be seen, especially in the L-H section.
}
\label{NaSi6_bands}
\end{center}
\end{figure}


\begin{figure}[ht!]
\begin{center}
\includegraphics[width=0.7\textwidth, angle=0]{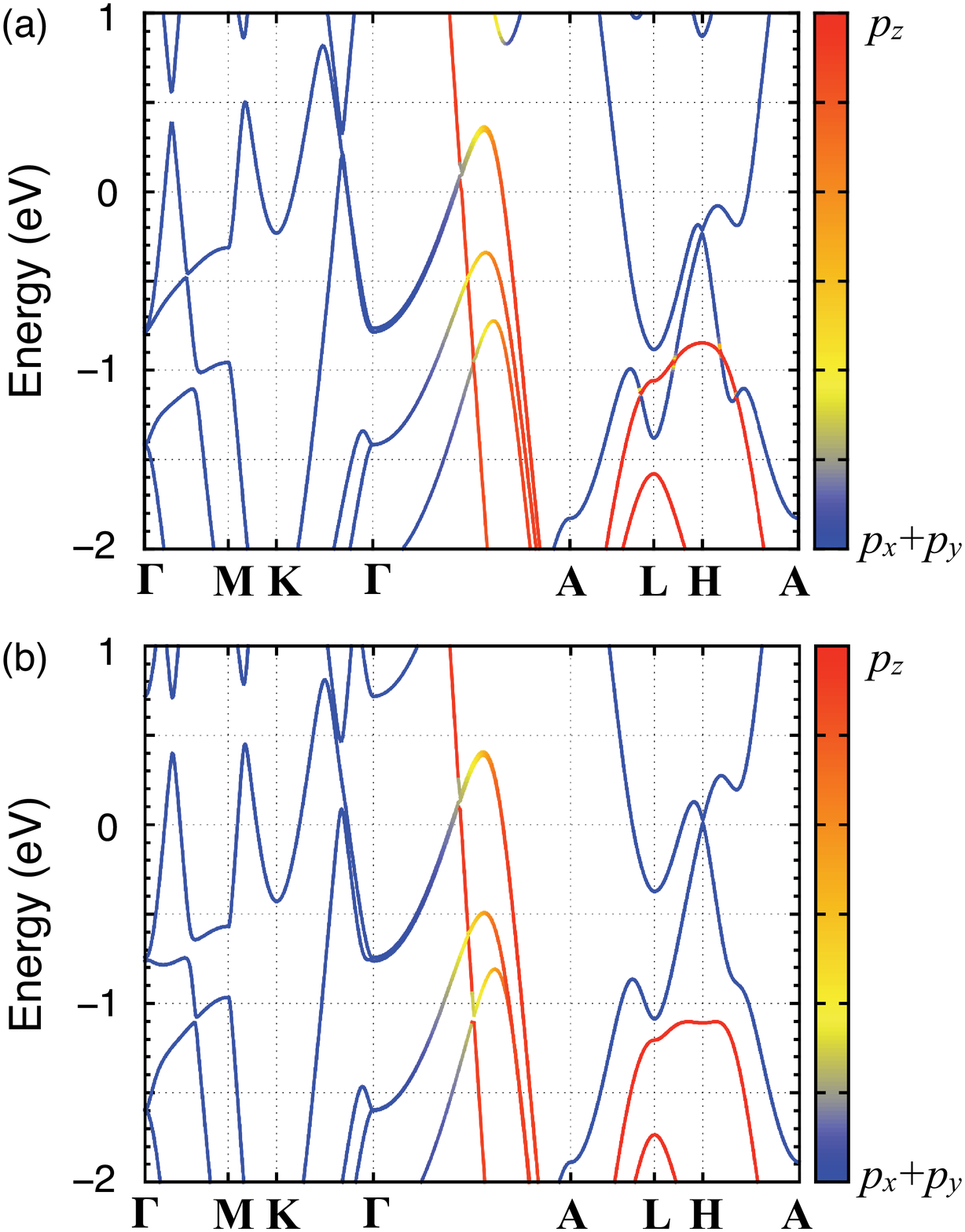}
\caption{Electronic band structure obtained from first-principles calculations using DFT.
In the presence of SOC, band structures of $P6/m$-Si$_6$ at (a) 0 GPa and (b) 15 GPa
and their projections on the the Si $p$ character are shown.
According to the relative weight of in-plane Si $p$ character ($p_x + p_y$) 
and out-of-plane Si $p$ character ($p_z$)
the band colors were encoded as shown in the legend. 
The situation of band inversion that occurs with changes in external pressure ($> 11.5$ GPa) 
can be seen,  especially in the L-H-A interval.
}
\label{Si6_p}
\end{center}
\end{figure}

\section{$\mathbb{Z}_2$ topological invariant calculations}

\begin{table}[h!]
\begin{ruledtabular}
\begin{center}
\caption{ Parities at time-reversal invariant momentum points of $P6/m$-Si$_6$ and $P6/m$-NaSi$_6$ at 0 GPa and 15 GPa.
}
\label{parities}
\begin{tabular}{c c c c c c}
&		& \multicolumn{2}{c}{$P6/m$-Si$_6$}	& \multicolumn{2}{c}{$P6/m$-NaSi$_6$} \\
\cline{3-4} \cline{5-6}
& {\bf k}-point	& 0 GPa	      & 15 GPa	 & 0 GPa	& 15 GPa \\
\hline
\rm{$\Gamma$}	& (0,0,0)	      & $+$	& $+$		& $-$	& $-$	\\
\rm{M$_1$}		& (0.5, 0, 0)	  & $+$	& $+$		& $-$	& $-$	\\
\rm{M$_2$}		& (0, 0.5, 0)	  & $+$	& $+$		& $-$	& $-$	\\
\rm{A}			& (0, 0, 0.5)	  & $+$	& $+$		& $-$	& $-$	\\
\rm{M$_3$}		& (0.5, 0.5, 0)	  & $+$	& $+$		& $-$	& $-$	\\
\rm{L$_1$}		& (0.5, 0, 0.5)	  & $-$	& $+$		& $+$	& $-$	\\
\rm{L$_2$}		& (0, 0.5, 0.5)	  & $-$	& $+$		& $+$	& $-$	\\
\rm{L$_3$}		& (0.5, 0.5, 0.5) & $-$	& $+$		& $+$	& $-$	\\
\end{tabular}
\end{center}
\end{ruledtabular}
\end{table}

\begin{figure}[ht!]
\begin{center}
\includegraphics[width=0.7\textwidth, angle=0]{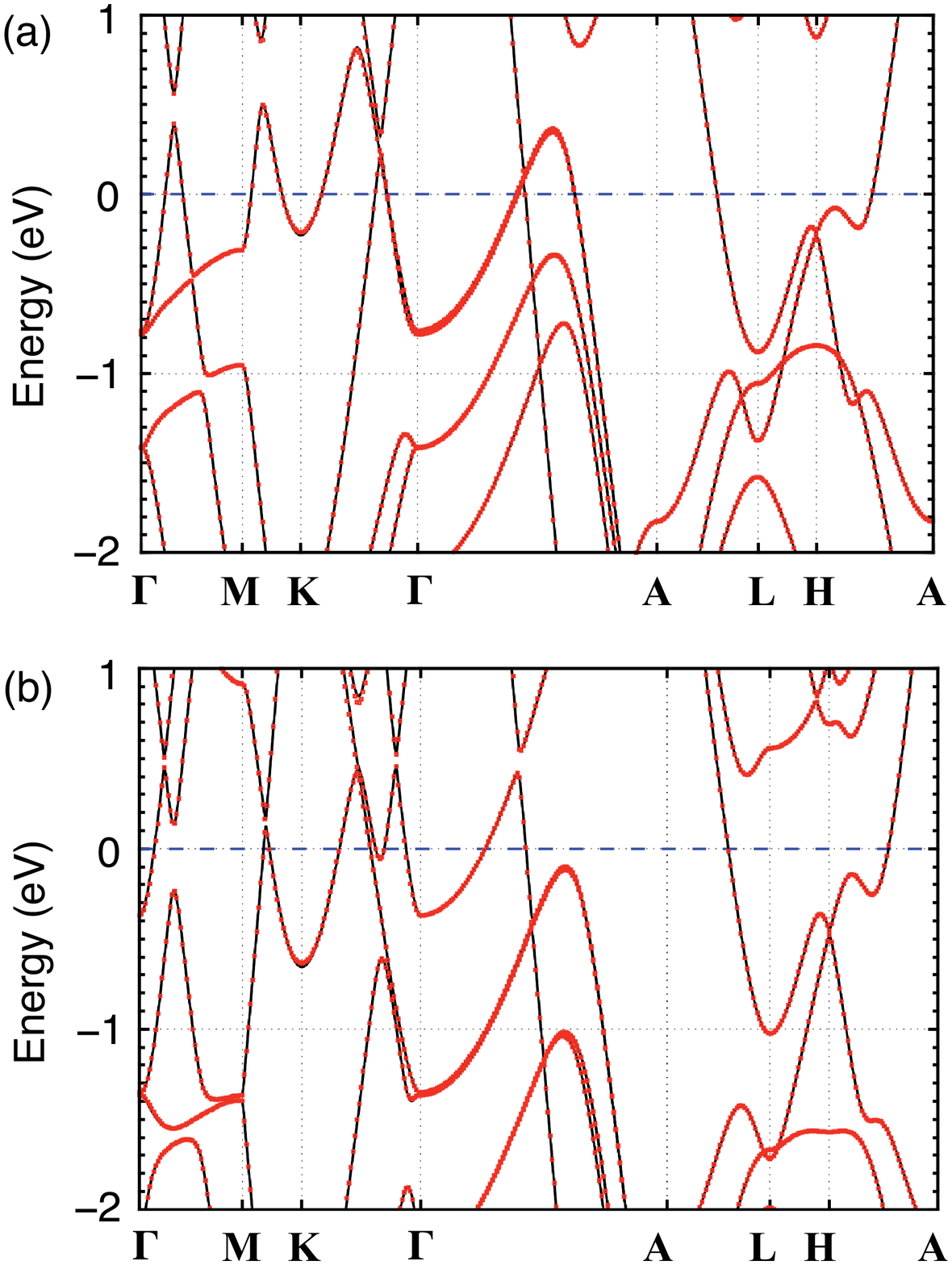}
\caption{ First-principles bands (solid line) 
and Wannier bands (red filled circles) 
of $P6/m$-Si$_6$ at  (a) 0 GPa and (b) 15 GPa.
In the eV energy scale, there are no visible differences between two different bands.  
Spin-orbit coupling is included. 
}
\label{wann_bands}
\end{center}
\end{figure}

The topological phase of band structure can be classified by calculating topological invariant.
The hybrid Wannier charge center flow technique, which is equivalent to the Wilson loop method.
To investigate the topological nature of the electronic states, 
we obtained maximally localized Wannier functions using the \textsf{WANNIER90} code \cite{wannier20} and performed hybrid Wannier charge center flow calculations.
With this Wannier model, 
the symmetry properties and $\mathbb{Z}_2$ topological invariant could be directly computed. 
$\mathbb{Z}_2$ topological invariants were calculated by observing hybrid Wannier charge center flows. The topological surface states were calculated using the Green's function approach, 
as implemented in the \textsf{WannierTools} code \cite{WU2017}.

Space group no. 175 ($P6/m$, hexagonal) has inversion symmetry.
Because of the existence of inversion symmetry in both $P6/m$-Si$_6$ and $P6/m$-NaSi$_6$, 
the Fu-Kane parity criterion can be used to easily calculate the $\mathbb{Z}_2$ topological invariants.
We were able to obtain the same $\mathbb{Z}_2$ topological invariants for both $P6/m$-Si$_6$ and $P6/m$-NaSi$_6$
by using two different methods, 
observing hybrid Wannier charge center flows 
and investigating Fu-Kane parity criterion.
$\nu_0$ is the strong topological index. 
$\nu^{\prime}_n$ ($n=1, 2, 3$) are weak topological indices.
Table \ref{nu_si6} show the example for $P6/m$-Si$_6$ at 0 GPa.
From the hybrid Wannier charge center flow calculations, we calculated $\nu^{\prime}_n$ ($n=1,2,3$).
$\nu_0$ is obtained by 
\begin{equation} \tag{S1}
\nu_0 = \nu_1+\nu^{\prime}_1 = \nu_2+ \nu^{\prime}_2 = \nu_3+ \nu^{\prime}_3 ~~~{\rm mod ~2} ~,
\end{equation}
and thus we have $(\nu_0; \nu^{\prime}_1 \nu^{\prime}_2 \nu^{\prime}_3) = (1; 0 0 1)$.
Therefore, $P6/m$-Si$_6$ at 0 GPa has a strong topological nature.
Similarly, we find that $P6/m$-NaSi$_6$ at 0 GPa has a strong topological nature.

\begin{table}
\begin{ruledtabular}
\begin{center}
\caption{ $\nu_0$ and $\nu^{\prime}_n$ ($n=1, 2, 3$) for $P6/m$-Si$_6$
obtained from hybrid Wannier charge center flow calculations, 
and correspond $\{k_n\}$ points and {\bf k}-planes. 
}
\label{nu_si6}
\begin{tabular}{c c c c}
$k_n$ & $k$-plane & $\nu$ \\
\hline
 $k_1$=0.0	& $k_2$-$k_3$ plane	& $\nu_1$=1	\\
 $k_1$=0.5	& $k_2$-$k_3$ plane	& $\nu^{\prime}_1$=0	\\
 $k_2$=0.0	& $k_1$-$k_3$ plane	& $\nu_2$=1	\\
 $k_2$=0.5	& $k_1$-$k_3$ plane	& $\nu^{\prime}_2$=0	\\
 $k_3$=0.0	& $k_1$-$k_2$ plane	& $\nu_3$=0	\\
 $k_3$=0.5	& $k_1$-$k_2$ plane	& $\nu^{\prime}_3$=1	\\
\end{tabular}
\end{center}
\end{ruledtabular}
\end{table}


\begin{table}[ht!]
\begin{ruledtabular}
\begin{center}
\caption{ Irreducible representation (irrep), point group symmetries, 
and their eigenvalues of bands at L point.
$C_2$ is two-fold rotation operator with respect to the $z$-axis.
$I$ and $\sigma_h$ are two-fold inversion and horizontal mirror plane operators, respectively. 
The highest occupied band is 12th band. }
\begin{tabular}{c c c c c}
band \# &	irrep		& $C_2$	& $\sigma_h$	& $I$ 		\\
\hline
1	& $B_u$	& $-1$	& $+1$	& $-1$	\\
2	& $A_g$	& $+1$	& $+1$	& $+1$	\\
3	& $B_g$	& $-1$	& $-1$	& $+1$	\\
4	& $B_u$	& $-1$	& $+1$	& $-1$	\\
5	& $A_u$	& $+1$	& $-1$	& $-1$	\\
6	& $A_g$	& $+1$	& $+1$	& $+1$	\\
7	& $B_g$	& $-1$	& $-1$	& $+1$	\\
8	& $A_u$	& $+1$	& $-1$	& $-1$	\\
9	& $B_u$	& $-1$	& $+1$	& $-1$	\\
10	& $A_g$	& $+1$	& $+1$	& $+1$	\\
11	& $A_u$	& $+1$	& $-1$	& $-1$	\\
12	& $B_u$	& $-1$	& $+1$	& $-1$	\\
\hline
13	& $B_g$	& $-1$	& $-1$	& $+1$	\\
\end{tabular}
\end{center}
\end{ruledtabular}
\label{symm_at_L}
\end{table}

\section{ S\lowercase{i}-S\lowercase{i} bond lengths in the slab geometry }

\begin{figure}[ht!]
\begin{center}
\includegraphics[width=0.6\textwidth, angle=0]{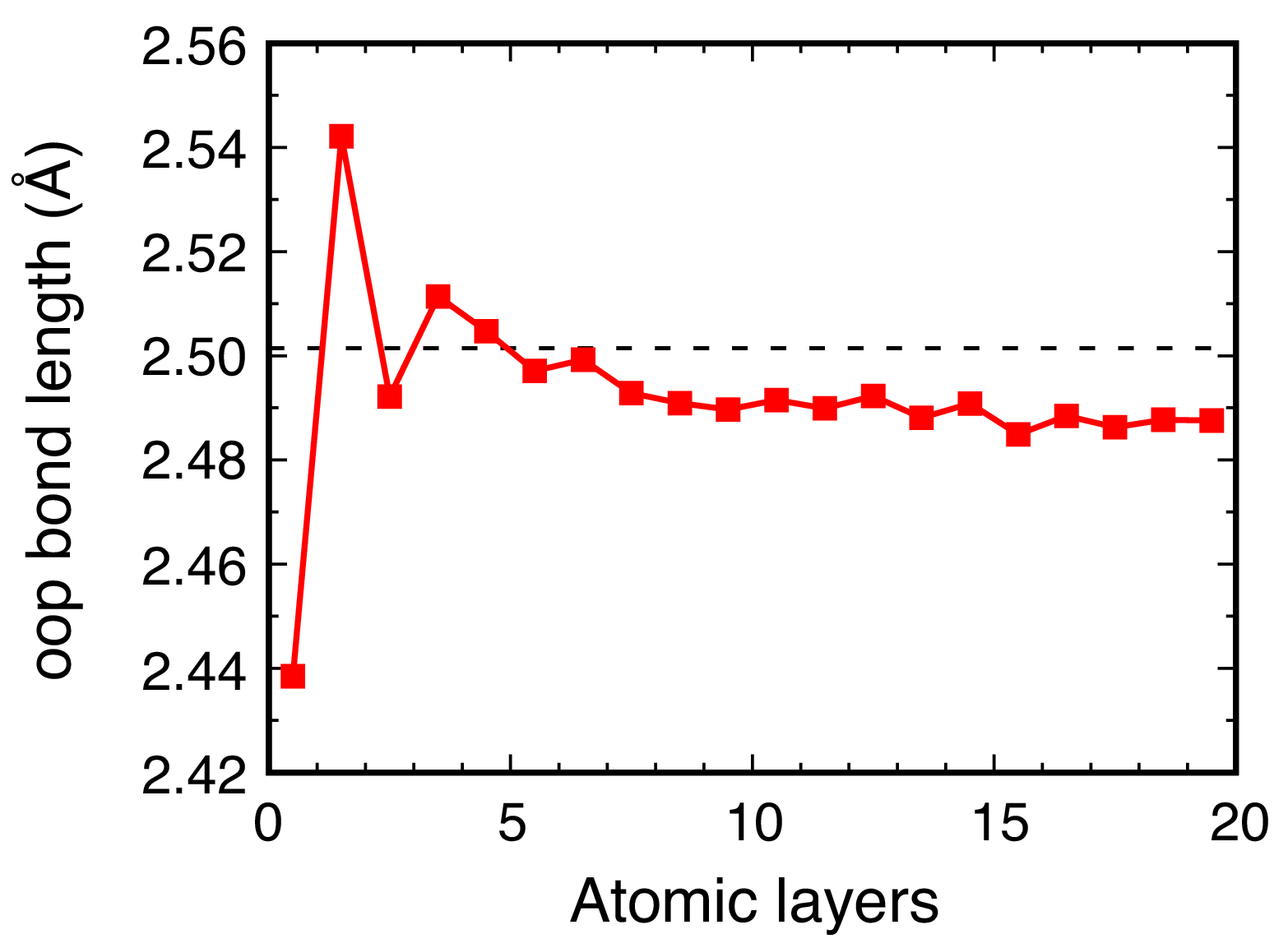}
\caption{ Si-Si bond lengths, formed in the $z$ direction 
also denoted by out-of-plane (oop bond lengths) from the hexagonal plane, 
in the relaxed 40-layer $P6/m$-Si$_6$ slab are shown. 
Position of the surface along the out-of-plane, is set to be zero. 
Dashed line represents the Si-Si bond length in the unrelaxed slab.
}
\label{bond_mod}
\end{center}
\end{figure}

\section{Fermi surfaces for $P6/m$-S\lowercase{i}$_6$ and $P6/m$-N\lowercase{a}S\lowercase{i}$_6$}
\begin{figure}[ht!]
\begin{center}
\includegraphics[width=0.98\textwidth, angle=0]{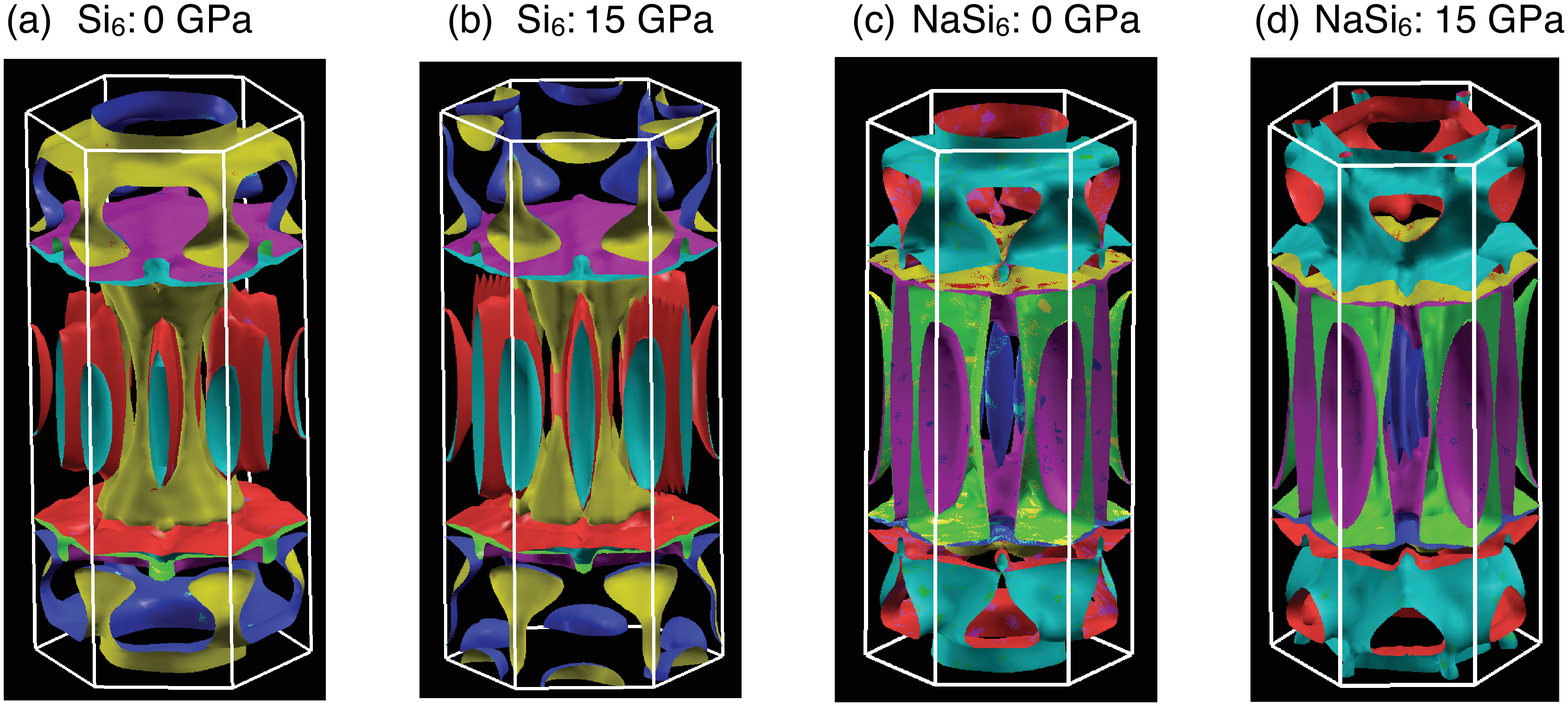}
\caption{Electronic band structure obtained from first-principles calculations using DFT.
Fermi surfaces of (a) $P6/m$-Si$_6$ at 0 GPa, (b) $P6/m$-Si$_6$ at 15 GPa,
(c) $P6/m$-NaSi$_6$ at 0 GPa, and (d) $P6/m$-NaSi$_6$ at 150 GPa
are shown, respectively. 
The Fermi surface is the surface of constant energy in the first BZ 
which separates occupied from unoccupied electron states at zero temperature.
}
\label{fermisurf2}
\end{center}
\end{figure}



\bibliography{myref}